\documentclass[reprint,aps]{revtex4}
\usepackage{amsfonts,amssymb,amsmath,epsfig,epic,color}
\usepackage{graphicx}
\usepackage{dcolumn}
\usepackage{bm}

\newcommand{\be}{\begin{equation}}
\newcommand{\ee}{\end{equation}}
\newcommand{\beqs}{\begin{eqnarray}}
\newcommand{\eeqs}{\end{eqnarray}}
\DeclareMathOperator{\slim}{s-lim}

\begin{document}
\title{Nondegeneracy of the Ground State for Nonrelativistic Lee Model}

\author{Fatih Erman}
\email{fatih.erman@gmail.com} \affiliation{Department of
Mathematics, Izmir Institute of Technology, Urla,
35430, Izmir, Turkey}%

\author{Berkin Malko\c{c}}
\affiliation{Physics Engineering Department,
Istanbul Technical University, 34469, Maslak, Istanbul, Turkey}

\author{O. Teoman Turgut}
\email{turgutte@boun.edu.tr} \affiliation{Department of Physics,
Bo\u{g}azi\c{c}i University, Bebek, 34342,
Istanbul, Turkey}

\begin{abstract} 
In the present work, we first briefly sketch the construction of the nonrelativistic Lee model on Riemannian manifolds, introduced in our previous works. In this  approach, the renormalized resolvent of the system is  expressed in terms of a well-defined operator, called the  principal operator, so as to obtain  a finite formulation. Then, we show that the ground state of the nonrelativistic Lee model on  compact Riemannian manifolds is nondegenerate using the explicit expression of the principal operator that we obtained. This is achieved by combining heat kernel methods with positivity improving semi-group approach and then applying these tools  directly  to the principal operator, rather than the Hamiltonian, without using cut-offs. 
\end{abstract}

\maketitle

\section{Introduction} \label{Introduction}

The Lee model \cite{lee} is a well-known nontrivial toy model for describing
the interaction of two relativistic chargeless spinless fermions, called $V$ and $N$ (e.g. ``nucleons"), with a scalar chargeless boson, called $\theta$ (e.g. ``pion"), through the only allowable process:
\be V \leftrightharpoons N + \theta \;, \ee
where the energies of the nucleons are assumed to be independent of their momenta (i.e., the recoil of the nucleon is neglected). There are two conserved quantities $Q_1$ and $Q_2$, which correspond to the total number of fermions and the difference between the number of $N$ and $\theta$ particles, respectively. Due to absence of antiparticles in the model (crossing symmetry is violated), and the above conserved quantities, the Fock space splits into a direct sum of invariant Hilbert spaces spanned by the restricted basis vectors labeled by the eigenvalues of $Q_1$ and $Q_2$. This makes the model  solvable since once $Q_1$ and $Q_2$ are fixed only a finite number of particles couples to each other. In addition to that, it is one of the first models where the coupling constant, mass and wave function renormalization can be carried out in a nonperturbative way.  An extensive discussion of this  model can be found in the textbook of Schweber \cite{Schweber}.

Following the idea of the original Lee model, its several variations have been studied. One simplified version is obtained by regarding the fermions so heavy that  their momenta are completely  neglected and the bosons are assumed to be nonrelativistic (i.e., their energy momentum relation satisfies $E={p^2 \over 2m} +m$ in the nonrelativistic approximation). In this model, only an additive renormalization
of the mass difference of the localized (or static) fermion states  is sufficient and it is performed in a closed form. This static  version of the Lee model has been discussed for the first two boson sectors in Henley and Thirring's book \cite{Henley Thirring} in detail. An extension of this model onto the Riemannian manifolds has been discussed in \cite{lee model 2 dimension, lee model 3 dimension} and the present  work is a continuation of those works for understanding  the ground state structure  of the system in depth. 

In the literature, there is a  great deal of  work devoted to some pathologies of Lee model, which appears when the renormalized coupling constant is greater than some critical value, and their possible  resolutions are worked out. Essentially, this happens because of a new state of the $V$-particle having an energy that is below the mass of the ``normal" V-particle \cite{Pauli, bender, jones}. However, our interest here is to focus on the nonperturbative nature of the model as such and the issue of introducing a physical V-particle will not  be  addressed.

Our approach in the present work  largely follows  the unpublished work by Rajeev 
\cite{rajeevbound}. There, he  introduced a new nonperturbative formulation of
renormalization for some simple nonrelativistic quantum mechanical and quantum field theoretical models where the particles interact with a point source. In his approach, the basic idea is  to work out  the resolvent of the Hamiltonian in the Fock space formalism and thereby identifying the divergent part explicitly. After removing the divergent part of the problem via this new renormalization procedure, a finite formulation of the model is accomplished. The resolvent contains  the  inverse of a new operator, called principal operator $\Phi(E)$. In this way, the whole renormalization procedure is carried out in the resolvent formalism without worrying about the self-adjoint extension of the Hamiltonian and its domain issues directly, which is essential for describing point interactions rigorously \cite{Albeverio}. Since the resolvent includes all the information about the spectrum of the problem, the bound states can then be found from its poles.  Point interactions in quantum mechanics, Lee model, and a model where the bosons are interacting through the two-body point potentials are studied from this point of view in \cite{rajeevbound}.

Following the ideas developed in \cite{rajeevbound}, we have extended the nonrelativistic Lee  model defined on flat spaces onto  Riemannian manifolds with the help of heat kernel techniques \cite{lee model 3 dimension, lee model 2 dimension}.  The full resolvent of the renormalized model is expressed in terms of the principal operator without an explicit expression of the renormalized Hamiltonian. In other words, we have obtained the analog of Krein's formula for the resolvent. Furthermore, we have proved  in \cite{existence} that there exists a densely defined  self-adjoint Hamiltonian operator corresponding to the resolvent we found and that the  ground state energy is bounded from below. The principal operator approach for Lee model on Riemannian manifolds can also be extended to interacting bosons on a two dimensional manifold \cite{bosons on manifolds}.

In the study of the bound states in quantum mechanics and quantum field theory, one of the main problems is to prove the uniqueness of the ground state. In general, the proof is not so trivial (non-uniqueness could also lead to interesting physics). Nevertheless, for sufficiently regular potentials, the proof is given by Courant and Hilbert \cite{Courant Hilbert} by \textit{implicitly} assuming all the regularity conditions on the potential in the context of Sturm-Liouville systems. A modern and a more rigorous treatment of the nondegeneracy for  the ground state  has actually been first given by Glimm and Jaffe \cite{Glimm Jaffe}  in the context of quantum field theory for self-interacting bosons ($\lambda (\phi^{4})_{2}$ interaction) by using the infinite dimensional extension of Perron - Frobenius theorem \cite{Horn} and positivity arguments from semi-group theory \cite{reedsimonv4}. 
Then, the applicability of these techniques in proving the nondegeneracy of the ground states to the Schr\"{o}dinger operators has been developed by Simon and H\"{o}egh-Krohn \cite{Simon Hoegh Krohn} for some regular class of potentials and this new modern version of the proof is also given in the textbooks \cite{reedsimonv4, Galindo Quantum mechanics1}.

The problem is mathematically formulated in the following way: For a given self-adjoint Hamiltonian $H$ describing the dynamics of the quantum mechanical or quantum field theoretical system, one must prove that Hamiltonian is bounded from below and the infimum of the spectrum is an eigenvalue. The eigenvector corresponding to that eigenvalue is then called the ground state. This completes the proof for the existence of the ground state. Then, one must prove that the eigenspace corresponding to that eigenvalue is one dimensional, i.e., the ground state is nondegenerate (unique up to complex multiples). The proof is essentially based on the Hilbert space generalization of the Perron-Frobenius theorem developed for nonnegative matrices. If we have a quantum mechanical model, then the Hilbert space is $L^2$ over some measure space. If we have a purely bosonic  quantum field theoretical model, then the positivity arguments are most naturally introduced in the tensor product space obtained from the coordinate representation, so called $Q$ space \cite{reedsimonv2, reedsimonv4}.  In this representation,  the concept of positivity becomes clear, in the present work, we will also be using the natural $L^2$-space on the manifold and the symmetrized tensor products thereof.
For positivity, in general it is easier to work with a bounded operator $e^{-tH}$ rather than $H$ itself. The largest eigenvalue of
$e^{-tH}$ will become $e^{-t E_0}$ if $E_0$ is the ground state energy for $H$. Then, assuming the Hamiltonian operator is self-adjoint and bounded from  below and  ground state corresponds to an eigenvector, the positivity improving property of the operator $e^{-tH}$ for all $t>0$ (which will be defined in Section \ref{Positivity and Nondegenaracy of the Ground State}) is  equivalent to the statement that the ground state energy $E_0$ of a self-adjoint Hamiltonian $H$, is non-degenerate (which is also equivalent to the associated eigenvector being  strictly positive).

After the inspiring work of Glimm and Jaffe \cite{Glimm 
Jaffe}, Gross \cite{Gross} and Faris \cite{Faris} extended it to the models involving fermions and bosons in the more abstract framework where the Hilbert space is not a standard $L^2$ space. Since then, the nondegeneracy of the ground state for several quantum field theory models, such as polaron models, spin-boson models, the van Hove model,
the Wigner-Weisskopf model, and non-relativistic quantum electrodynamics have been proved and discussed extensively from several point of views \cite{Sloan, Spohn, Arai Hirokawa1, Arai Hirokawa2, Bach Frohlich Sigal, Miyao, Abdesselam1, Abdesselam2}.

In the framework of our formulation for finitely many $N$ point delta interactions in two and three dimensional Riemannian manifolds \cite{point interactions on manifolds}, we have proved that the resolvent after the renormalization procedure is given by a kind of Krein's formula and expressed in terms of a $N \times N$ matrix $\Phi$, called principal matrix. The matrix $\Phi(E)$ includes all the information about the bound state spectrum, the values  of which is found by solving the equation $\Phi(E) A=0$. In other words, the zero modes of the principal matrix correspond to the bound state spectrum of the problem. Since we do not have an expression for the renormalized Hamiltonian, the proof for the nondegeneracy of the ground state can not be given in the same way as developed for the regular potentials \cite{reedsimonv4}. Nevertheless, the principal matrix allows us to prove the uniqueness of the ground state, via the Perron-Frobenius theorem applied directly to the principal matrix. Since the ground state eigenvector is expressed in terms of the eigenvector of the principal operator associated to its minimum eigenvalue and their degeneracies are equal, this proves the claim. 

In this work, we will generalize the arguments developed for the point interactions to prove the nondegeneracy of the ground state of the Lee model defined on Riemannian manifolds,  the construction of which were already established in our previous studies \cite{lee model 3 dimension, lee model 2 dimension} by extending the ideas introduced in \cite{point interactions on manifolds}. Although the basic idea in proving the nondegeneracy of the ground state in this model is similar to the one which we developed for point interactions, the proof requires the use of the positivity arguments.  The main difference between our method and the one given for other field theory models \cite{reedsimonv4} is that we have no formal expression of the Hamiltonian after our nonperturbative renormalization procedure.   
Hence, we can not apply the positivity arguments for the semi-group generated by the Hamiltonian operator, instead we will directly use them for the principal operator using the results given in \cite{existence} and then we will be able to prove the nondegeneracy of the ground state without using any cut-off.

The paper is organized as follows. In Section \ref{Summary of the Construction of Nonrelativistic Lee Model on Manifolds}, we will first present a very brief overview of the basic results for the construction of the nonrelativistic Lee model on $D=2,3$ dimensional Riemannian manifolds and then give a more detailed analysis on the ground state of the problem. Finally, we prove that the ground state wave function is strictly positive, so that ground state is nondegenerate for compact Riemannian manifolds with Ricci curvature bounded below. The proof for noncompact manifolds is technically much more challenging and exceeds the present skills of the authors.  We comment that compactness can be thought of as a kind of infrared regularization, from this point of view, such a restriction should not be an  essential handicap in understanding {\it only the ultraviolet complications} of these problems. 

In this paper, we will use the notations $\langle \cdot |\cdot \rangle$ or $\langle \cdot , \cdot  \rangle$ to denote the inner product and $|| \, | \cdot \rangle ||$ or $||\cdot||$ to denote the associated norms.

\section{Summary of the Construction of Nonrelativistic Lee Model on Manifolds} \label{Summary of the Construction of Nonrelativistic Lee Model on Manifolds}

In order to make our discussion reasonably self-contained,  first we  shortly give the important results of our approach for  the renormalization of the model presented  in \cite{lee model 3 dimension, lee model 2 dimension} and add some new comments about the ground state of the problem.

The regularized Hamiltonian in a $D$ ($2$ or $3$) dimensional Riemannian manifold
$(\mathcal{M},g)$ is formally given by
\begin{eqnarray}
H^{\epsilon} & = & H_0 + H_{I,\epsilon} = 1 \otimes \int_{\mathcal{M}}  
\phi^\dag_{g}(x)\left(-{1\over 2m}\nabla^2_g+m \right)\phi_g(x) \; d_{g}^{D} x
\cr & + & \mu (\epsilon) \left( {1-\sigma_3\over 2} \right) \otimes 1 +
\lambda \int_{\mathcal{M}} 
K_\epsilon(x,a)\left[\sigma_{-} \otimes \phi_g(x)+ \sigma_+ \otimes \phi^\dag_{g}(x) 
\right] \; d_{g}^{D} x \;, \label{regH}
\end{eqnarray}
where $\epsilon$ is a cut-off parameter (we use the units such that $\hbar=c=1$), $\lambda$ is the coupling constant,
and $\nabla^2_g$ is the Laplace - Beltrami operator. Also, $d^{D}_{g} x$ is the Riemannian volume element and $x,y$ refer to points on the manifold $ \mathcal{M}$. 
The Hilbert space is $\mathbb{C} \otimes \mathcal{F}_{b}$, where $\mathcal{F}_{b}$ is the bosonic Fock space. The function 
$ K_\epsilon(x,a)$ is the heat kernel on a Riemannian manifold
with metric structure $g$ and it converges to the Dirac delta
function $\delta_g (x,a)$ around the point $a$ on the manifold as
we take the limit $\epsilon \to 0^+$ \cite{Grigoryan Heat Book}. The creation and annihilation operators $\phi^{\dag}_{g}(x)$, $\phi_{g}(x)$ defined on $(\mathcal{M},g)$ obey the following canonical commutation relations
\be
[\phi_{g}(x),\phi^{\dag}_{g}(y)] = \delta_g (x,y) \;. \label{commutation}
\ee
Also, the matrices
$\sigma_{\pm}$ given in the Hamiltonian are the standard Pauli spin-flip matrices. Similar to the flat space case, the coefficient 
$\mu(\epsilon)$ denotes the bare mass difference between the
$V$ particle (neutron) and the $N$ particle (proton). The conserved charge $Q_2$ is 
\begin{equation} 
- \left({1+ \sigma_3 \over 2}\right) \otimes 1 + 1 \otimes \int_{\mathcal{M}} \phi^{\dag}_g(x) \phi_g(x) \; d_{g}^{D} x  \;,
\end{equation}
which makes  the model solvable. Since $\mathbb{C} \otimes \mathcal{F}_{b}= \mathcal{F}_{b} \oplus \mathcal{F}_{b}$, we can represent the regularized Hamiltonian as a
$2\times 2$ block matrix:
\begin{equation} H^{\epsilon} -E =
\left(%
\begin{array}{cc}
  H_0-E & \lambda \int_{\mathcal{M}}  K_\epsilon(x,a)\, \phi^\dag_g(x) \; d_{g}^{D} x \\
   \lambda \int_{\mathcal{M}}  K_\epsilon(x,a)\, \phi_g(x) \; d_{g}^{D} x & H_0- E
   +\mu(\epsilon)
\end{array}%
\right)\;.
\label{regH} \end{equation}
For given $Q_1=1$ and $Q_2=n$, the above Hamiltonian acts on the sector $\mathcal{F}_{n+1} \oplus \mathcal{F}_{n}$, where $\mathcal{F}_{n}$ stands for the symmetrized $n$ tensor product of the one particle Hilbert spaces. 
If we suppose that the regularized resolvent is of the following form
\begin{equation}
R^{\epsilon}(E)= (H^{\epsilon} -E)^{-1}= \left( \begin{array}{cc}
  a_\epsilon & b^\dagger_\epsilon \\
 b_\epsilon & d_\epsilon \\
\end{array}\right)^{-1} = \left(%
\begin{array}{cc}
  \alpha_\epsilon & \beta^{\dagger}_{\epsilon} \\
  \beta_{\epsilon} & \delta_{\epsilon} \\
\end{array}%
\right) \;,
\end{equation}
one can find $\alpha_\epsilon$, $\beta_{\epsilon}$, and $\delta_{\epsilon}$ in terms of $a_\epsilon$, $b_\epsilon$ and $d_\epsilon$ given in Eq. (\ref{regH}) in two apparently different but equivalent ways (see the appendix in \cite{rajeevbound} for the explicit computation)
\begin{eqnarray}
\alpha_\epsilon&=& (H_0-E)^{-1}+ (H_0-E)^{-1} \;
b^{\dag}_{\epsilon} \; \Phi_\epsilon^{-1} (E) \; b_\epsilon \;
(H_0-E)^{-1} \cr \beta_\epsilon &=&- \Phi^{-1}_\epsilon(E)\;
b_\epsilon \; (H_0-E)^{-1} \cr \delta_\epsilon
&=&\Phi^{-1}_\epsilon(E)\cr b_\epsilon &=&\lambda
\int_{\mathcal{M}} 
K_\epsilon(x,a)\,\phi_g(x) \; d_{g}^{D} x  \;,
\end{eqnarray}
where
\begin{equation}
\Phi_\epsilon(E)=H_0-E+\mu(\epsilon)-\lambda^2
\int_{\mathcal{M}^2} K_\epsilon(x,a) K_\epsilon(y,a)\; \phi_g(x)  (H_0-E)^{-1} \phi^\dagger_g(y) \; d_{g}^{D} x \, d_{g}^{D} y \label{phi_e}\;,
\end{equation}
called regularized \textit{principal operator}. By using the eigenfunction expansion of the creation and the annihilation operators with the commutation relations (\ref{commutation}), we find \cite{lee model 3 dimension} 
\beqs & &  (H_0-E)^{-1}
\phi^{\dag}_{g}(y) = \int_{\mathcal{M}}  \int_0^\infty \phi^{\dag}_{g}(y) e^{-t (H_0-E + m)} K_{t}(y,y') \; d t \; d_{g}^{D} y' \;,
\label{normalorderingcreation1}  \eeqs
and 
\beqs & & 
\phi_{g}(x) (H_0-E)^{-1} =
\int_{\mathcal{M}} \int_0^\infty  e^{-t(H_0-E + m)}  K_{t}(x,x')  \phi_{g}(x')  \; d t \;
d_{g}^{D} x' \;.
\label{normalorderingannihilation1}  \eeqs
If we
normal order the principal operator (\ref{phi_e}) using the above relations (\ref{normalorderingcreation1}) and (\ref{normalorderingannihilation1}),  its singular structure   becomes transparent 
\begin{eqnarray}
&\ & \Phi_\epsilon(E)= H_0-E-\lambda^2 \int_{\epsilon/2}^\infty
 \int_{\mathcal{M}^2}  K_{t}(x,a)K_{t}(y,a) \phi^{\dag}_{g}(x)
e^{-(t-\epsilon/2)(H_0+2m-E)}\phi_{g}(y)  \; d_{g}^{D} x \,
d_{g}^{D} y \; d t \cr &\ &
\hspace{3cm}    + \; 
 \mu(\epsilon) - \lambda^2  \int_\epsilon^\infty
K_{t}(a,a) \; e^{-(t-\epsilon)(H_0+m-E)}\; d t  \;. \label{phiepsilon}
\end{eqnarray}
(we warn the reader that there is a  typo in the corresponding  equation of this result  in \cite{existence, lee model 2 dimension}, which does not change the final expression of the principal operator). It is now easy to see that the last term in Eq. (\ref{phiepsilon}) is divergent due to the short time asymptotic expansion of the diagonal heat kernel given by
\be
K_t(x,x) \sim {1 \over (4 \pi t /2m)^{D/2}} \; \sum_{k=0}^{\infty} u_k(x,x) (t/2m)^k \;,
\ee
for every $x$ in any $D$ dimensional Riemannian manifold without boundary \cite{Gilkey} and the functions $u_k(x,x)$ are scalar polynomials in the curvature tensor of the manifold and its covariant derivatives at $x$.
This suggest that by choosing $\mu(\epsilon)$
\begin{eqnarray} \label{mueps}
\mu(\epsilon) = \mu + \lambda^2 \int_{\epsilon}^{\infty}
 K_{t} (a,a)\, e^{-t(m-\mu)} \; d t \;,
\end{eqnarray}
where $\mu$ being the experimentally measured bound state energy of the composite
state which consists of a boson and the attractive heavy neutron
at the center $a$, and substituting it in (\ref{phiepsilon}) and then taking the limit $\epsilon\to 0^+$, we get the following finite expression
\begin{eqnarray} & & 
\Phi(E)=H_0-E+\mu+\lambda^2 \int_0^\infty 
K_{t}(a,a)\left[e^{-t(m-\mu)} -e^{-t(H_0+m-E)}\right] \; d t \cr & \ &
\hspace{3cm} -\lambda^2 \int_0^\infty 
\int_{\mathcal{M}^2} 
K_{t}(x,a)K_{t}(y,a) \, \phi^\dag_g(x)
e^{-t(H_0+2m-E)}\phi_g(y)  \; d_{g}^{D} x \, d_{g}^{D} y \; d t \;. \label{phi lee} \end{eqnarray}
Therefore, we have
a well-defined explicit formula for the full resolvent of the
Hamiltonian in terms of the inverse of the principal operator
$\Phi(E)$ and the free resolvent, namely
\be
R(E) = \left(%
\begin{array}{cc}
  \alpha(E) & \beta^{\dagger}(E) \\
  \beta(E) & \delta(E) \\
\end{array}%
\right) \;, \label{resolvent}
\ee
where 
\begin{eqnarray}
\alpha(E) &=& (H_0-E)^{-1}+ (H_0-E)^{-1} \;
\phi^{\dag}(a) \; \Phi^{-1} (E) \; \phi(a) \;
(H_0-E)^{-1} \cr \beta(E) &=& - \Phi^{-1}(E)\;
\phi(a) \; (H_0-E)^{-1} \cr \delta(E)
&=&\Phi^{-1}(E) \;. \end{eqnarray}

\section{Ground State}
\label{Ground State}

For compact and connected Riemannian manifolds, the spectral theorem for Laplace - Beltrami operator \cite{chavel} states that :

\textit{There exists a complete orthonormal system of $C^\infty$
eigenfunctions $f_{\sigma}(x)$ in $L^2$ 
\begin{equation}
-\nabla_g^2 f_{\sigma}(x)=\sigma f_{\sigma}(x) \;, 
\end{equation}
with purely discrete spectrum $\{\sigma_l\} = \{0 = \sigma_0
\leq \sigma_1 \leq \sigma_2 \leq \dots \leq \sigma_N \leq \ldots\}$, where $\sigma_l$ tending
to infinity as $l \rightarrow \infty$ and each eigenvalue has
finite multiplicity.}

This implies that the spectrum of the second quantized free Hamiltonian $H_0$ in the given sector must also be discrete.  Since our main interest here is in the bound state spectrum, we will require from now on that $m > \mu$, which guarantees that heavy particles do not decay. Moreover, although we have no explicit expression for the renormalized Hamiltonian after our renormalization procedure, we have proved  in \cite{existence} that there exists a densely defined self-adjoint Hamiltonian operator associated with the renormalized resolvent. Hence, the spectrum of the Hamiltonian must be real.

The poles of the resolvent $R(E)$ in (\ref{resolvent}) give the bound state spectrum of the Hamiltonian operator. \textit{For compact manifolds}, it turns out that the ground state of the system in the given sector must only come from the roots of the equation
\be \Phi(E) |\Psi \rangle =0 \;, \ee
since the free resolvent has no poles in the given sector (e.g., the spectrum of $H_0$ starts from $(n+1)m>nm +\mu$ in the $(n+1)$ boson sector). This is due to the fact that the ground state energy is below $n m + \mu$, which will be shown by a variational ansatz at the end of this section. In other words, the zero modes of the principal operator are responsible for the ground state of the system.

It is well known that the contour integral of the resolvent around any pole of it in the complex $E$ plane is the projection operator $\mathbb{P}_k$ onto
the associated eigenspace of the Hamiltonian, given by Riesz integral formula
\begin{equation} \label{projection resolvent}
\mathbb{P}_{k}= -{1 \over 2 \pi i} \oint_{\Gamma_{k}} R(E) \; d E \;,
\end{equation}
where $\Gamma_{k}$ is a small contour enclosing the isolated
eigenvalue $E_k$ \cite{reedsimonv4}. Let us first consider the first sector of the
resolvent
and choose the contour $\Gamma_0$, enclosing only the ground state energy
$E_{gr}$. Then, the contour integral gives the projection onto the eigenspace associated with $(n+1)$ boson sector, represented by $|\Psi_0^{(n+1)} \rangle \; \langle \Psi_0^{(n+1)}|$.

We have proved that the principal operator
$\Phi(E)$ is  a self-adjoint holomorphic family of type A in the sense of Kato
\cite{existence}, so that we can apply the generalized spectral theorem for 
its inverse, namely
\beqs \Phi^{-1}(E) &=&\sum_{k} {1 \over
\omega_k(E)} |\omega_k(E) \rangle \langle \omega_k(E) |  \;,  \eeqs
where $\omega_k(E)$ and
$|\omega_k(E) \rangle$
are the eigenvalues and the eigenvectors of the principal
operator, respectively. We have removed any possible continuous part of the resolution of our operator,  since as we will prove below in Sec. \ref{Positivity and Nondegenaracy of the Ground State},  {\it the principal operator $\Phi(E)$ only has a discrete spectrum}.  Let $\omega_0 (E)$ be the minimum eigenvalue of the principal operator for all $E$ by assuming it exists for the moment (which we will prove in Sec. \ref{Positivity and Nondegenaracy of the Ground State}). Then, Feynman-Hellman theorem applied to $\omega_0$ implies that (which is  valid even  for  degenerate states) the flow of $\omega_0$ with respect to $E$ is 
\beqs & & {\partial \omega_0\over \partial E} = \langle
\omega_0(E) | {\partial \Phi(E) \over \partial E} | \omega_0(E)
\rangle = - \Bigg( 1 + \lambda^2 \int_{0}^{\infty}  t \; K_t(a,a) \left|\left| e^{-{t \over
2}(H_0-\mu + E)}| \omega_0(E) \rangle \right| \right|^2 \; d t \cr & & \hspace{1cm} + \;
\lambda^2 \int_{0}^{\infty}  t \;  \left| \left|
e^{-{t \over 2} (H_0 - \mu + E)}\int_{\mathcal{M}}
K_t(x,a) \phi_g(x) | \omega_0(E) \rangle \; d_{g}^{D} x
\right|\right|^2 \; d t\Bigg)  \;. \label{derivative of eigenvalue
phi}\eeqs
From this equation, the positivity property of the heat kernel $K_t(x,y)>0$ for all $x,y \in \mathcal{M}$ and $t>0$ leads to the following inequality
\be
{\partial \omega_0\over \partial E} <0 \;.
\label{flow of omega} \ee
As a consequence of this important fact (\ref{flow of omega}), the 
ground state energy must correspond to the zero of the minimum eigenvalue
$\omega_0(E)$. Expanding 
$\omega_0(E)$ near the ground state energy $E_{gr}$
\begin{equation}
\omega_0(E) = \omega_0(E_{gr}) + (E- E_{gr}) {\partial \omega_0(E)
\over
\partial E}\bigg|_{E_{gr}}+ \cdots  = (E- E_{gr}) {\partial \omega_0(E)
\over
\partial E}\bigg|_{E_{gr}}+ \cdots  \;,
\end{equation}
and using the residue theorem in (\ref{projection
resolvent}), we obtain
\beqs |\Psi_0^{(n+1)} \rangle \;  \langle \Psi_0^{(n+1)} | = (H_{0}-E_{gr})^{-1} \phi^{\dag}(a) \left(- {\partial
\omega_0(E)\over
\partial E}|_{E_{gr}}  \right)^{-1}
|\omega_0(E_{gr}) \rangle \langle \omega_0(E_{gr}) | \phi(a)
(H_{0}-E_{gr})^{-1} \;, \label{residue resolvent} \eeqs
where
\be | \omega_0(E_{gr}) \rangle = \int_{\mathcal{M}^{n}}
\psi_{0}(x_1,\cdots, x_{n}) |x_1 \cdots x_{n} \rangle \; d_{g}^{D}x_1 \cdots d_{g}^{D}x_{n}  \;. \ee
is a generic form of the ground state of the principal operator. Then, by repeated applications of the formula (\ref{normalorderingcreation1}) and (\ref{normalorderingannihilation1}), we can
shift all the creation operators $\phi^{\dag}_{g}(x)$ to the leftmost
\beqs & &  (H_0-E)^{-1}
\phi^{\dag}_{g}(a)\phi^{\dag}_{g}(x_1)\cdots\phi^{\dag}_{g}(x_{n})
= \int_{\mathcal{M}^{n+1}}  \phi^{\dag}_{g}(y_1)\cdots
\phi^{\dag}_{g}(y_{n+1}) \int_0^\infty e^{-t (H_0-E + (n+1)m)} \nonumber \\
\cr  & & \hspace{4cm} \times    \,
K_{t}(y_1,a) K_{t}(y_2,x_1)\cdots K_{t}(y_{n+1},x_{n}) \; d t \; d_{g}^{D} y_1 \cdots
d_{g}^{D} y_{n+1} \;,
\label{normalorderingcreation}  \eeqs
and all the annihilation operators $\phi_{g}(x)$ to the rightmost
\beqs & & 
\phi_{g}(a)\phi_{g}(x_1)\cdots\phi_{g}(x_{n}) (H_0-E)^{-1} =
\int_{\mathcal{M}^{n+1}} \int_0^\infty  \; e^{-t(H_0-E + (n+1) m)}  K_{t}(y_1,a) K_{t}(y_2,x_1)\cdots
K_{t}(y_{n+1},x_{n}) \nonumber \\
\cr & & \hspace{5cm} \times \,  \; \phi_{g}(y_1)\cdots \phi_{g}(y_{n+1}) \; d t \;  d_{g}^{D} y_1 \cdots
d_{g}^{D} y_{n+1}  \;.
\label{normalorderingannihilation}  \eeqs
in Eq. (\ref{residue
resolvent}), so that we can read off the ground state vector $|\Psi_0^{(n+1)} \rangle$ in the $(n+1)$ boson sector
\beqs & & |\Psi_{0}^{(n+1)} \rangle = \int_{\mathcal{M}^{n+1}}
\Psi_{0}(y_1, \ldots, y_{n+1}) |y_1\cdots y_{n+1} \rangle \; d_{g}^{D} y_1 \cdots d_{g}^{D} y_{n+1}  \cr & & =
\int_{\mathcal{M}^{n+1}}  \int_{\mathcal{M}^{n}}
{1
\over (n+1)} \sum_{\sigma \in (1 \cdots (n+1))} \int_0^\infty 
\; e^{-t((n+1)m-E_{gr})}
K_{t}(y_{\sigma(1)},a) K_{t}(y_{\sigma(2)},x_1)\cdots
K_{t}(y_{\sigma(n+1)},x_{n}) \nonumber \\
\cr & &  \times \; \psi_{0}(x_1,\cdots,x_{n})
\left(- {\partial \omega_0(E)\over
\partial E}|_{E_{gr}}  \right)^{-1/2}
|y_{\sigma(1)}\cdots y_{\sigma(n+1)} \rangle \; d t \; d_{g}^{D} y_1 \cdots
d_{g}^{D} y_{n+1}  \; d_{g}^{D} x_1 \cdots d_{g}^{D} x_{n} \label{psi0_1} \;. \eeqs
Here the sum runs over all cyclic permutations $\sigma$ of
$(123\ldots (n+1))$ since the wave function $\Psi_0$ must be symmetric. Similarly, we can compute the ground state in the $n$ boson sector from the residue integral, and obtain
\beqs
|\Psi_0^{(n)} \rangle = \left(- {\partial \omega_0(E)\over
\partial E}|_{E_{gr}}  \right)^{-1/2} \; \int_{\mathcal{M}^{n}}
\psi_{0}(y_1, \ldots, y_{n}) |y_1\cdots y_{n} \rangle \; d_{g}^{D} y_1 \cdots d_{g}^{D} y_{n} \label{psi_0_2} \;.
\eeqs
Hence, the ground state of the system is given in the following form
\begin{equation} |\Psi_0 \rangle =
\left(%
\begin{array}{cc}
  |\Psi_0^{(n+1)} \rangle  \\
  |\Psi_0^{(n)} \rangle 
\end{array}%
\right)\;,
   \end{equation}
where $|\Psi_0^{(n+1)} \rangle$ and 
  $|\Psi_0^{(n)} \rangle$ are given in (\ref{psi0_1}) and (\ref{psi_0_2}), respectively. 
Notice that if the right hand side of the contour integral of the resolvent only includes one dimensional projection operators, so is the left hand side, which will be of fundamental importance in our proof for the nondegeneracy of the ground state. 

We will now demonstrate that the zero of the minimum eigenvalue of the principal operator $\omega_0(E)$ occurs for a value below $nm+\mu$, hence it is enough to study this operator family for  the sector  $E \leq nm+\mu$. 
Let us now make the following variational ansatz
\begin{equation} |\omega_{var}\rangle= {1\over V({\cal M})^{n/2}}  {1 \over \sqrt{n!}} \; \int_{{\cal M}^n} \phi_{g}^\dagger(y_1) \ldots \phi_{g}^\dagger(y_n)| 0 \rangle \; d_{g}^{D} y_1 \ldots d_{g}^{D} y_n \;, 
\end{equation}
and choose $E=E_{var}=nm+\mu$. Here $V(\mathcal{M})$ denotes the volume of the compact Riemannian manifold $\mathcal{M}$. By the variational principle, the lowest eigenvalue of the principal operator $\Phi(E_{var})$ satisfies the inequality
\begin{eqnarray}
\omega_0(E_{var})\leq \langle \omega_{var}| \Phi(E_{var})|\omega_{var}\rangle \;.
\end{eqnarray}
We note that $ \langle \omega_{var} | H_0|\omega_{var} \rangle=nm$ and $e^{-t(H_0+m-E_{var})}|\omega_{var}\rangle=e^{-t(m-\mu)}|\omega_{var}\rangle$. This implies that 
\begin{equation}
       \omega_0(E_{var})\leq \langle\omega_{var}| U(E_{var})|\omega_{var}\rangle \;. 
\end{equation}
Then, we get
\begin{equation}
  \omega_0(E_{var})\leq -\lambda^2 \int_0^\infty \langle\phi_g(K_t(.,a))\omega_{var}| e^{-t(H_0+2m-E_{var})}|\phi_g(K_t(.,a))\omega_{var}\rangle \; dt \;, 
\end{equation}
where we have defined the following compact notation, which will be useful in Sec. \ref{Positivity and Nondegenaracy of the Ground State}
\begin{equation}
 \phi_g(K_t(.,a))=\int _{\cal M}  K_t(y,a) \phi_g(y) \; d^D y \;. 
\end{equation}  
Since
\begin{eqnarray}
|\phi_g(K_t(.,a))\omega_{var}\rangle &= & {\sqrt{n} \over V({\cal M})^{n/2}} {1 \over \sqrt{(n-1)!}} \int_{\cal M}  K_t (x,a)  \int_{{\cal M}^{n-1}} \phi^{\dagger}_{g} (y_1)...\phi^{\dagger}_{g} (y_{n-1})| 0 \rangle  \; d_{g}^{D}y_1 \ldots d_{g}^{D}y_{n-1} \; d_{g}^{D} x \nonumber\\
&=& {\sqrt{n}\over V({\cal M})^{n/2}} {1 \over \sqrt{(n-1)!}}\int_{{\cal M}^{n-1}} \phi^{\dagger}_{g}(y_1) \ldots \phi^{\dagger}_{g} (y_{n-1})| 0 \rangle \; d_{g}^{D} y_1 \ldots d_{g}^{D} y_{n-1}  \;,  
\end{eqnarray}
and the operator $e^{-t(H_0+2m-(nm+\mu))}$ acting on this wave function brings a multiplicative factor $e^{-t(m-\mu)}$, we have
\begin{equation}
       \omega_0(E_{var})\leq \langle\Psi_{var}| U(E_{var})|\Psi_{var}\rangle=-{n\lambda^2 \over (m-\mu) V(\cal M)} \;,
\end{equation}
after taking the inner product and integrating over $t$.
This expression is strictly negative, so that
in order to find the solution $\omega_0(E_{gr})=0$ we must reduce $E$ below $E_{var}$ due to (\ref{flow of omega}).
As a result, the ground state energy indeed is below $nm+\mu$ and corresponds to the zero of $\omega_0$, as claimed.

We have also proved in \cite{lee model 3 dimension, lee model 2 dimension, existence} that the Hamiltonian is bounded from below and this lower bound is given by
\begin{equation} E_{*} = nm+\mu -(n \lambda^2 C)^{{2 \over 4-D}} \label{E bound}\;,
\end{equation}
where $C$ is a positive constant.

\section{Positivity and Nondegenaracy of the Ground State} \label{Positivity and Nondegenaracy of the Ground State}

In this section we will show that the ground state wave function of the nonrelativistic Lee model defined in two and three dimensional compact Riemannian manifold can be chosen as positive, and as a consequence of this it is nondegenarate. The key idea behind this  is to study some positivity properties of the semi-group $e^{-t \Phi}$ generated by the principal operator $\Phi$ rather than the Hamiltonian, which is usually  the standard method used  to prove the nondegeneracy of the ground states for some field theory and quantum models \cite{reedsimonv4}. We know from Sec. \ref{Ground State} that all the information about the ground state of the system is hidden in the principal operator, that is, the solutions of the zeros of its minimum eigenvalue  give the ground state energy. Therefore, it is natural to study the positivity properties of the semi-group $e^{-t \Phi}$ instead of $e^{-t H}$.

Let us first remind some terminology of the positivity  \cite{reedsimonv4} in Hilbert space $\mathcal{H}= L^2(\mathcal{M}, d \mu)$. A function $\psi \in L^2(\mathcal{M}, d \mu)$ is called \textit{positive} if it is nonnegative almost everywhere and is not the zero function ($\psi \geq 0$). It is called \textit{strictly positive} if $\psi>0$ almost everywhere.  A bounded operator $A$ on $L^2(\mathcal{M}, d \mu)$ is called \textit{positivity preserving} if $A\psi$ is positive whenever $\psi$ is positive. In order to show the nondegeneracy of the ground state, we need a slightly stronger positivity property: $A$ is called \textit{positivity improving} if $A\psi$ is strictly positive whenever $\psi$ is positive.  A bounded operator $A$ on $L^2(\mathcal{M}, d \mu)$ is positivity improving if and only if $(f, A g) >0$ for all positive functions $f,g \in L^2(\mathcal{M}, d \mu)$. We may use this to show that the Laplace -Beltrami operator on a manifold generates a positivity improving semi-group.

The notion of positivity can be extended onto symmetric (bosonic) Fock spaces $\mathcal{F}_b(\mathcal{H}) =  \oplus_{n=0}^{\infty} S_n  \otimes^n  L^2({\cal M}, d \mu)$, that is, if the real-valued functions on one-particle Hilbert spaces $L^2(\mathcal{M}, d \mu)$ are positive, then it implies that  
the function on the symmetric Fock space, constructed from the one-particle Hilbert spaces is also positive. For a fermionic system, this would not be true due to the minus signs under permutations.

\subsection{Existence of the Ground State}
\label{Existence of the Ground State}

In general, if we are given a self-adjoint Hamiltonian $H$, which is bounded from below, and the infimum of the spectrum is an eigenvalue, then we say that the ground state exists for the given model Hamiltonian $H$. However, we do not have an explicit expression for the Hamiltonian yet, we have shown that there exists a self-adjoint Hamiltonian associated to the renormalized resolvent and this Hamiltonian is bounded from below \cite{existence}. 
It suffices to prove that the infimum of the spectrum is an eigenvalue for the proof of the existence of the ground state. We will prove this as follows: 
We first prove that the infimum of the spectrum of the principal operator $\Phi$ is an eigenvalue, that is,
\be
\inf \sigma(\Phi(E)) = \omega_0 (E) \;,
\ee
for all $E\leq nm+\mu$. Then it follows that the infimum of the spectrum of the Hamiltonian is an eigenvalue due to the fact that the first eigenvalue $\omega_0$ of the principal operator is a monotonically decreasing function of $E$ (see Eq. (\ref{flow of omega})) and the unique solution for the zero of  this eigenvalue  corresponds to the ground state of the model. 

For simplicity, let us decompose the principal operator $\Phi(E)$ into three parts:
\be
\Phi(E)=K_0(E) + K_1(E) + U(E)  \;,
\ee
where
\beqs
K_0(E) &=& 
H_0-E+\mu \cr 
K_1 (E)&=&\lambda^2\int_0^\infty  K_t(a,a) \left(e^{-(m-\mu)t}-e^{-(H_0+m-E)}\right) \; d t
\cr U(E) & = & - \lambda^2 \int_0^\infty 
\int_{\mathcal{M}^2} 
\; K_{t}(x,a)K_{t}(y,a) \, \phi^\dag_g(x)
e^{-t(H_0+2m-E)}\phi_g(y)  \; d_{g}^{D} x \, d_{g}^{D} y \; d  t \;.
\eeqs
As a consequence of the compactness of the manifold, the essential spectrum of $H_0$ (or $K_0$) is empty, that is,
\be
\sigma_{ess}(K_0) = \sigma_{ess}(H_0) = \varnothing \;.  
\ee
Since the kinetic part $K_1$ is a function of $H_0$, we expect that the essential spectrum of $K_0 + K_1$ is also empty. In order to show this explicitly, we will first prove that $K_1$  is a relatively compact perturbation of the kinetic part $H_0$ by showing that $(K_{0}-z)^{-1} \; K_1$ for some $z$ in the resolvent set of $K_0$ is a trace-class operator when raised to a certain power, say $4n$ (every trace class operator is compact \cite{reedsimonv1}). We first remark that 
\begin{equation}
 tr_{\mathcal{F}_{n}} \left(e^{-t H_0} \right) = \left[tr_{\mathcal{H}} \left(e^{-t \left(-{1 \over 2m} \nabla_{g}^{2} +m\right)} \right) \right]^{n} = e^{- t n m} \left(\int_{\mathcal{M}}  K_t(x,x)  \; d_{g}^{D} x \right)^n
\label{trace} \;, \end{equation}
where $tr_{\mathcal{F}_{n}}$ stands for the trace over the symmetrized $n$ tensor product of one particle Hilbert spaces. For  compact manifolds, Ricci curvature is bounded from below, i.e., $Ric(\cdot, \cdot) \geq -\kappa g (\cdot, \cdot)$, assuming $\kappa \geq 0$ to cover the most general case, and the upper bound of the  diagonal heat kernel for any $t>0$ and $x \in \mathcal{M}$  is given by
\begin{equation}
K_t(a,a)\leq {1 \over V({\cal M})} + {C\over (t/2m)^{D/2}} \;, \label{heat upper bound}
\end{equation}
where $C$ depends on $\kappa$ and the diameter, and  the volume of the manifold \cite{Wang}.
Then, the operator $e^{-t H_0}$ is trace class, hence a compact operator for all $t \in (0, \infty)$. In our proof these facts will be essential.

Without loss of generality, let $z=0$ for simplicity. Then using the integral representation of the operator $K_0^{-1} K_1$ 
\begin{equation}
\int_{0}^{\infty} \int_{0}^{1} t \; K_t(a,a) e^{-t u (H_0 + \mu -E)} e^{- t m} \; d u  \; d t \;,
\end{equation}
the trace of $4n$ th power of this operator becomes
\begin{eqnarray} & & 
\int_{\mathbb{R}_{+}^{4n}} \int_{0}^{1}  \int_{0}^{1}  \; \ldots \int_{0}^{1}  t_1 \ldots t_{4n} \; tr_{\mathcal{H}^{(n)}_{s}} \left(e^{-(t_1 u_1 + \ldots + t_{4n} u_{4n}) (H_0 +\mu-E)} \right) K_{t_1}(a,a) \ldots K_{t_{4n}}(a,a) \cr & & \hspace{5cm} \times \;  e^{-(t_1 + \ldots + t_{4n}) m} \; d u_1  \; d u_2 \ldots d u_{4n} \; d t_1 \ldots d t_{4n} \;. 
\end{eqnarray}
After scaling the variables $t_i u_i$ to $t_i$, we get
\begin{eqnarray} & & 
\int_{\mathbb{R}_{+}^{4n}} \int_{0}^{1}  \int_{0}^{1}  \; \ldots \int_{0}^{1}  \;  {t_1 \ldots t_{4n} \over u_{1}^{2} \ldots u_{4n}^{2}} \; tr_{\mathcal{H}^{(n)}_{s}} \left(e^{-(t_1 + \ldots + t_{4n}) (H_0 +\mu-E)} \right) K_{t_1/u_1}(a,a) \ldots K_{t_{4n}/u_{4n}}(a,a) \cr & & \hspace{5cm} \times \;  e^{-(t_1/u_1 + \ldots + t_{4n}/u_{4n}) m} \; d u_1 \; d u_2 \ldots d u_{4n} \;  d t_1 \ldots d t_{4n} \;. 
\end{eqnarray}
Using Eq. (\ref{trace}), the above result becomes 
\begin{eqnarray} & & 
\int_{\mathbb{R}_{+}^{4n}}  \int_{0}^{1} \int_{0}^{1}  \ldots \int_{0}^{1}  \;  {t_1 \ldots t_{4n}  \over u_{1}^{2} \ldots u_{4n}^{2}} \; e^{-(t_1 + \ldots + t_{4n}) (n m +\mu-E)} \left( \int_{\mathcal{M}} \; K_{t_1+\ldots +t_{4n}}(x,x) \; d_{g}^{D} x \right)^{n} \;   \cr & & \hspace{2cm} \times \; K_{t_1/u_1}(a,a) \ldots K_{t_{4n}/u_{4n}}(a,a) \; e^{-(t_1/u_1 + \ldots + t_{4n}/u_{4n}) m} \; d u_1  \; d u_2 \ldots d u_{4n} \; d t_1 \ldots d t_{4n} \;  \label{trace2} \;.
\end{eqnarray}
From the upper bound of the diagonal heat kernel given in (\ref{heat upper bound}), an upper bound of Eq. (\ref{trace2}) is obtained 
\begin{eqnarray} & & \left( V^{n}(\mathcal{M}) C^{5n} (2m)^{5nD/2}\right)
\int_{\mathbb{R}_{+}^{4n}}   \int_{0}^{1}  \; \int_{0}^{1}  \; \ldots \int_{0}^{1}   {t_1 \ldots t_{4n} \over u_{1}^{2} \ldots u_{4n}^{2}} \; e^{-(t_1 + \ldots + t_{4n}) (n m +\mu-E)}  \cr & & \hspace{2cm} \times \; { (u_1 \ldots u_{4n})^{D/2} \over (t_1 \ldots t_{4n})^{D/2}}\; {1 \over (t_1 + \ldots + t_{4n})^{nD/2}}  e^{-(t_1/u_1 + \ldots + t_{4n}/u_{4n}) m}  \; d u_1 \; d u_2  \ldots  d u_{4n}  \; d t_1 \ldots d t_{4n} \label{trace3} \;,
\end{eqnarray}
by taking account of only the most singular terms, that is   we
disregard contributions coming from the volume term. As one can check, those terms that we dropped behave much better. Using the arithmetic-mean inequality, 
\begin{equation}
{1 \over t_1 + \ldots + t_{4n}} \leq {1 \over (4n)^{nD/2} (t_1 \ldots t_{4n})^{D/8}} \;, 
\end{equation}
Eq. (\ref{trace3}) is less than
\begin{eqnarray} & & \left(V^{n}(\mathcal{M}) C^{5n} (2m)^{5nD/2}\right) \left(\int_{0}^{1}
 \int_{0}^{\infty} u^{{D \over 2}-2} {e^{-t (n m +\mu-E +{m \over u})}  \over t^{7/8}} \; d t \; d u  \right)^{4n} \;. 
\end{eqnarray} 
Evaluating the $t$-integral and using the fact that $E \leq nm +\mu$, the upper bound to Eq. (\ref{trace3}) becomes  
\begin{eqnarray}\left(V^{n}(\mathcal{M}) C^{5n} (2m)^{5nD/2}\right) \left(\int_{0}^{1} 
u^{{D \over 2}-2} ({u \over m})^{1/8}\; { \Gamma(1/8)  } \; d u \right)^{4n}  \;,  \end{eqnarray}
which is finite, i.e.,
\begin{equation}
tr_{\mathcal{F}_{n}}([K_{0}^{-1}K_1]^{4n}) < \infty \;. 
\end{equation}
Hence the essential spectra of the $K_0$ and $K_0 + K_1$ must coincide due to classical Weyl's theorem \cite{reedsimonv4}
\be
\sigma_{ess}(K_0+K_1) = \sigma_{ess}(K_0) = \varnothing \;. 
\ee
We should now prove that the same is true for the potential part, that is $U(E)$ is a relatively compact perturbation of $K_0$, hence by Weyl's theorem
again their essential spectra must coincide \cite{reedsimonv4}, that is, 
\be \sigma_{ess}(K_0 + U)=
\sigma_{ess}(K_0)  = \varnothing  \;,
\ee
and this means that
\be
\sigma_{ess}(\Phi) = \varnothing \;.
\ee
In other words, everything in the spectrum of the principal operator is an eigenvalue.

We will now show explicitly that the operator $K_0^{-1} U$ can be approximated by a sequence of finite rank operators in the norm topology (which is sufficient to establish compactness). For that purpose, we choose the following basis,
\begin{equation}
| \varphi_{\sigma_1} \ldots \varphi_{\sigma_n} \rangle = {1 \over \sqrt{n!} \sqrt{\Pi_{i=1} n_{i}!}}  \sum_{\mathcal{P}} \mathcal{P} |\varphi_{\sigma_1} \ldots \varphi_{\sigma_n} ) \;, 
\end{equation}
where the sum runs over all possible permutations for the $n$-tuple $(1,2,\ldots,n)$ and $|\varphi_{\sigma_1} \ldots \varphi_{\sigma_n} ) $ is given in terms of the one-particle eigenstates, i.e., $|\varphi^{(1)}_{\sigma_1} \rangle |\varphi^{(2)}_{\sigma_2} \rangle \ldots | \varphi^{(n)}_{\sigma_n} \rangle $ (the upper indices refer to the particle label). Here, the permutation operator $\mathcal{P}$ is assumed to act on the particle indices.  Then, any $n$-particle state can be expanded in terms of this orthonormal basis.
\begin{equation}
    |\psi^{(n)}\rangle=\sum_{\sigma_1, \sigma_2,...,\sigma_n} \psi(\sigma_1, \sigma_2,...,\sigma_n)|\varphi_{\sigma_1},...,\varphi_{\sigma_n} \rangle \;.
    \end{equation}
Hence, when we write $\psi(\sigma_1,...,\sigma_n)$, whenever there are coincidences of these labels, the appropriate combinatoric factor is taken into account. Thus,
\begin{equation}
\sum_{\sigma_1,...,\sigma_n} |\psi(\sigma_1,...,\sigma_n)|^2=1 \;. 
\end{equation}
Let us now write down $\hat U(E)=[H_0-E+m]^{-1} U(E)$ in an eigenfunction expansion using the basis we refer above. An easy computation shows that,
\begin{eqnarray} 
          \hat U(E) | \psi^{(n)} \rangle &=& \lambda^2 \sum_{\sigma';\sigma_1 \sigma_2, ..., \sigma_n}{1\over n} \sum_{(12...n)} \int_0^\infty  \int_0^1  f_{\sigma'}(a) \; t \; e^{-t u({\sigma' \over 2m}+m-\mu)}  e^{-t(\sum_i {\sigma_i \over 2m}+(n+1)m-E)}  \cr
          &  & \hspace{2cm} \times \; f_{\sigma_1}(a)\psi(\sigma', \sigma_2,...\sigma_n) | \varphi_{\sigma_{1}}, \varphi_{\sigma_{2}}, \ldots, \varphi_{\sigma_{n}} \rangle \; d u \; d t 
\;, \end{eqnarray}
here the symbol $\sum_{(12...n)}$ refers to sum over cyclic permutations for the symmetrization and the matrix elements in this basis can easily be read.  Now we introduce the finite rank truncations of this operator,
simply by cutting-off at the $N$th eigenvalue for each block,
so that we have its finite dimensional approximation,
\begin{eqnarray} 
          \hat U_{N}(E)  |\psi^{(n)} \rangle  &=& \lambda^2 \sum_{[\sigma';\sigma_1 \sigma_2, ..., \sigma_n]\leq \sigma_N}{1\over n} \sum_{(12...n)} \int_0^\infty \int_0^1  f_{\sigma'}(a) \; s \; e^{-t u({\sigma' \over 2m}+m-\mu)}  e^{-t(\sum_i {\sigma_i \over 2m} +(n+1)m-E)} \cr
          &\ & \ \ \times \; f_{\sigma_1}(a)\psi(\sigma', \sigma_2,...\sigma_n) | \varphi_{\sigma_{1}}, \varphi_{\sigma_{2}}, \ldots, \varphi_{\sigma_{n}} \rangle \; du \; dt \;,
 \end{eqnarray}
where the symbol underneat reflects the fact that all the sums over the eigenvalues are upto the $N$th eigenvalue $\sigma_N$. We will show that $U_N$ strongly converges to $U$, that is, 
\begin{equation}
           || U(E)-U_{N}(E)||\to 0 \;, 
\end{equation}
as $N\to \infty$. The difference will have various blocks, so that we may represent these operators as block sums as follows,
\begin{equation}
\sum_{\sigma'>\sigma_N} \sum_{\sigma_1,...,\sigma_n} \oplus_{i=1}^{n} \sum_{\sigma' <\sigma_N} \sum_{\sigma_i > \sigma_N} \sum_{[\sigma_1,...,\sigma_{i-1}]<\sigma_N} \sum_{\sigma_{i+1},...,\sigma_n}  
\;. \end{equation}
We will now estimate the norm of the each term in the block sum after applying the norm inequality for each term represented by this block splitting. Since removing the restrictions in the indices, after we take the absolute values in the norms, increases the value of the sum, we will instead estimate norm of the following sum 
\begin{equation}
\sum_{\sigma'>\sigma_N} \sum_{\sigma_1,...,\sigma_n} \oplus_{i}  \sum_{\sigma_i>\sigma_N}\sum_{\sigma_1,...\hat \sigma_{i},...,\sigma_n} \sum_{\sigma'} \;, \end{equation}
which provides an upper bound. The index $\hat \sigma_i $ means that the sum over $\sigma_i$ is omitted. The norm square of the first sum turns out to be, 
\begin{eqnarray}
          &\ &  \lambda^4 \int_{0}^{\infty} \int_{0}^{\infty} \int_{0}^{1} \int_{0}^{1} \;   \sum_{\sigma_1,...,\sigma_n}\left( \sum_{\sigma'> \sigma_N} f_{\sigma'}(a)\psi(\sigma', \sigma_{(2)},...,\sigma_{(n)}) \; e^{-{t_1u_1\sigma' \over 2m}} \right) \cr
           &\ &  \hspace{2cm} \times \; \left(\sum_{\sigma^{''}> \sigma_N}
  f_{\sigma^{''}}(a) \psi(\sigma^{''},\sigma_{(2)},...,\sigma_{(n)}) \; e^{-{t_2 u_2\sigma^{''} \over 2m}}\right)   |f_{\sigma_{(1)}}(a)|^2 \; t_1 \; t_2 \; 
  e^{-(t_1u_1+t_2u_2)(m-\mu)} \cr & & \hspace{5cm} \times \; e^{-(t_1+t_2)(\sum_{i} {\sigma_{i} \over 2m} +(n+1) m-E)} \;du_2 \; du_1 \; d t_2  \; d t_1 \label{normsquare1}
\;, \end{eqnarray}
where the parenthesis over the indices $\sigma_{(i)}$ refer to the cyclic permutations. For the term coming from the action of the Hamiltonian, which gives $\sum_i \sigma_{i}$, if we only keep  the index which gives $\sigma_{(1)}$,  that becomes an upper bound. Moreover, the upper bound for the sums above inside the bracket can be easily  found as 
\begin{eqnarray}
     &\ & \Big| \sum_{\sigma > \sigma_N}  f_{\sigma}(a) \psi(\sigma, \sigma_{(2)},...,\sigma_{(n)})e^{-{t u \sigma \over 2m}} \Big| \leq   
   e^{-t u\sigma_N/4m} \Big| \sum_{\sigma> \sigma_N}
  f_{\sigma}(a) \psi(\sigma, \sigma_{(2)},...,\sigma_{(n)})e^{-t u\sigma/4m}\Big| \cr
  &\ & \leq e^{-t u\sigma_N/4m} \sum_{\sigma} \Big|
  f_{\sigma}(a) \psi(\sigma, \sigma_{(2)},...,\sigma_{(n)}) e^{-t u\sigma/4m}\Big| \cr 
    &\ & \leq  e^{-t u\sigma_N/4m} \Big[ \sum_{\sigma} |f_\sigma(a)|^2 e^{-t u\sigma/2m}\Big]^{1/2}\Big[\sum_{\sigma}        
    |\psi(\sigma,\sigma_{(2)}...,\sigma_{(n)}|^2\Big]^{1/2}\;,
\end{eqnarray}
where we have applied the Cauchy-Schwarz inequality in the last line. The sum $\sum_{\sigma} |f_\sigma(a)|^2 e^{-t u\sigma/2m}$ is the eigenfunction expansion of the  heat kernels $K_{t u}(a,a)$. We also note that the two sums over the wave function combine to give the norm of the wave function,
 the left-over index $\sigma_{(1)}$ again combines with $|f_{\sigma_{(1)}}(a)|^2$ and $e^{-(t_1+t_2)\sigma_{(1)}}$ to give another heat kernel. As a result we obtain the following upper bound for Eq. (\ref{normsquare1}),
\begin{eqnarray}
     &\ &  \lambda^4 \int_{0}^{\infty}  \int_{0}^{\infty} \int_{0}^{1} \int_{0}^{1}  \;   K_{t_1+t_2} (a,a) K_{t_1u_1}^{1/2}(a,a) K_{t_2 u_2}^{1/2}(a,a) \; t_1  t_2 \; \cr 
&\ & \hspace{2cm} \times \;   e^{-(t_1u_1+t_2u_2)\sigma_N/4m} e^{-(t_1u_1+t_2 u_2)(m-\mu)} e^{-(t_1+t_2)(m-E)}|| \psi||^2  \;  du_2  \; du_1 \; d t_2 \; d t_1 \;.
\end{eqnarray}  
Due to the upper bound of the diagonal heat kernel (\ref{heat upper bound}), the above expression is bounded above by
\begin{eqnarray}
     &\ &  (2m)^{D} \lambda^4 C^2 \int_{0}^{\infty} \int_{0}^{\infty}  \int_{0}^{1}  \int_{0}^{1}  \;  {1\over (t_1+t_2)^{D/2}}{1\over (t_1u_1)^{D/4}}{1\over (t_2u_2)^{D/4}} \; t_1 t_2 \cr 
&\ & \hspace{2cm} \times \;  \; e^{-(t_1u_1+t_2u_2)\sigma_N/4m}e^{-(t_1u_1+t_2u_2)(m-\mu)}e^{-(t_1+t_2)(m-E)}||\psi||^2  \;  du_2  \; du_1  \; d t_2  \; d t_1
\;. \end{eqnarray}
Now we scale the $t_1, t_2$ variables by $u_1, u_2$ respectively to get, after simplifications,
\begin{eqnarray}
&\ &   (2m)^{D} \lambda^2 C^2 \int_{0}^{\infty} \int_{0}^{\infty} \int_{0}^{1} \int_{0}^{1}  \;    
{t_1^{1-{D \over 4}} t_2^{1-{D \over 4}} \over (t_1 u_2+t_2 u_1)^{D/2} (u_1 u_2)^{{(4-D) \over 2}}}  \; e^{-(t_1+t_2)\sigma_N/4m} e^{-(t_1+t_2)(m-\mu)}  \cr & & \hspace{5cm} \times \; e^{-(t_1/u_1+t_2/u_2)(m-E)} ||\psi||^2 \;  du_2 \; du_1  \; d t_2 \; d t_1 \label{su} \;.
\end{eqnarray}
Using the arithmetic-mean inequality, 
\begin{equation}
{1\over (t_1 u_2+t_2 u_1)^{D/2}} \leq {1 \over 2^{D/2}} {1\over (t_1 t_2 u_1 u_2)^{D/4}} \;,
\end{equation}
we have decoupled the terms, and obtain the following upper bound to the norm in the first term
\begin{eqnarray}
&\ &   {(2m)^{D/2} \lambda^2 C ||\psi|| \over 2^{D/4}} \Bigg[ \int_{0}^{\infty}  \;    
t^{1-{D \over 2}}  \; e^{-t({\sigma_N \over 4m} +m-E)}  \Bigg( \int_{0}^{1} {e^{-(t/u)(m-E)} \over u^{2-{D \over 4}}} \;  du \Bigg) \; dt \Bigg] \label{su2} \;.
\end{eqnarray}
Since 
\begin{equation}
\int_{0}^{1} {e^{-(t/u)(m-E)} \over u^{2-{D \over 4}}} \;  du \leq \int_{0}^{\infty}  {e^{-(t/u)(m-E)} \over u^{2-{D \over 4}}} \;  du = \Gamma \left(1-{D \over 4} \right)  \; (m-E)^{{D \over 4}-1} t^{{D \over 4}-1} \;,
\end{equation}
the total result (\ref{su2}) is bounded from above by 
\begin{equation} {(2m)^{D/2} \lambda^2 C ||\psi|| \over 2^{D/4}} 
 \Gamma^2 \left(1-{D \over 4} \right) \;  (m-E)^{{D \over 4}-1} (\sigma_N)^{{D \over 4}-1},
\end{equation}
which goes to $0$ as $N\to\infty$ Note that here, we are using the important  fact about the eigenvalues, that they are of finite multiplicities, hence there is no infinite subsequence which remains bounded as we let $N\to \infty$.
Let us now consider the other sum (we have $n$-identical such terms),
\begin{eqnarray}
          &\ &  \int_{0}^{\infty}  \int_{0}^{\infty} \int_{0}^{1} \int_{0}^{1}  \sum_{\sigma_i>\sigma_N} \sum_{\sigma_1,...,\hat \sigma_i,...,\sigma_n}\left[ \sum_{\sigma'} f_{\sigma'}(a)\psi(\sigma', \sigma_{(2)},...,\sigma_{(n)})e^{-{t_1u_1\sigma' \over 2m}} \right] |f_{\sigma_{(1)}}(a)|^2 \cr
           &\ &   \hskip-0.5cm \times \; \left[\sum_{\sigma^{''}}
  f_{\sigma^{''}}(a) \psi(\sigma^{''},\sigma_{(2)},...,\sigma_{(n)})e^{-{t_2u_2\sigma^{''} \over 2m}}\right]    
  e^{-(t_1u_1+t_2u_2)(m-\mu)} e^{-(t_1+t_2)(\sum_{i} \sigma_{i} +(n+1)m-E)} du_2 \; du_1 \; d t_2 \; d t_1
  \;. \end{eqnarray}
We have two cases, as a result of symmetrization: one is that the restricted index shows up inside the wave function $\psi$, or it remains outside thereby it becomes the index of the eigenfunction $f_{\sigma_{(i)}}(a)$. If it is inside the wave function, we replace the last exponential sum by $e^{-(t_1+t_2)(\sigma_N/4m+(n+1)m-E)}$, (factor of $2$ is for convenience only). If it comes with the eigenfunction,  we split the eigenvalue part  into two equal pieces and replace the first one by  
$e^{-(t_1+t_2)(\sigma_N/4m)}$ and keep the remaining piece inside the sum to combine with the eigenfunction again. After this replacements,  we remove the restriction on the sum in both cases.
As a result,  by applying a Cauchy-Schwarz inequality to the mixed expression with $\sigma', \sigma^{''}$-terms, we find, 
\begin{eqnarray}
&\ & \leq \int_{0}^{\infty} \int_{0}^{\infty}  \int_{0}^{1} \int_{0}^{1}  e^{-(t_1\sigma_N/4m+t_2\sigma_N/4m)} \; K_{t_1u_1}^{1/2} (a,a) K_{t_2u_2}^{1/2} (a,a) \; 
e^{-(t_1+t_2)((n+1)m-E)}  \cr & & \hspace{4cm} \times \; K_{(t_1+t_2)/2}(a,a) \; du_2 \; du_1 \; d t_2 \; d t_1  \;.
\end{eqnarray}
If we now use the heat kernel estimates (\ref{heat upper bound}) and again use the most singular part with the arithmetic - mean inequality, the integrals become decoupled, so that the norm itself becomes smaller than 
\begin{equation}
  {C \lambda^2 (2m)^{D/2} \over 2^{D/4}} \int_{0}^{\infty} t^{1-{D \over 2}} e^{-t(\sigma_N/4m+(n+1)m-E)} \; d t \;  \int_{0}^{1} {1 \over u^{D/4}}  \; d u \; ||\psi|| \;.
  \end{equation}
Thus each one of these terms (there are $n$ of them) will go to zero since,
\begin{equation}
     \leq {4 C \lambda^2 (2m)^{D/2} \over 2^{D/4} (D-4)}\Gamma \left(2-{D \over 2}\right) \left(\sigma_N/4m +(n+1)m-E \right)^{{D \over 2} -2} ||\psi||\to 0  \ \ {\rm as }\quad N\to \infty \;. 
     \end{equation}
This implies that the bottom of the spectrum of the principal operator is indeed an eigenvalue, say $\omega_0(E)$, whether it is above the free part or not is of no concern, that is,
\be
\inf \sigma(\Phi) = \omega_0(E) \;,
\ee
for all $E\leq nm+\mu$. Hence, the infimum of the spectrum of the Hamiltonian associated with the renormalized resolvent is an eigenvalue, which completes the proof of the existence of  ground state wavefunction as a normalizable  state.

\subsection{Nondegeneracy of the Ground State}
\label{Uniqueness of the Ground State}

In order to prove the nondegeneracy of the ground state, we need the following theorem \cite{reedsimonv4} applied to the principal operator

\textit{Let $\Phi$ be a self-adjoint operator that is bounded from below. Suppose that $e^{-t \Phi}$ is positivity preserving for all $t>0$ and $\omega_0(E)= \inf \sigma(\Phi)$ is an eigenvalue. Then the following are equivalent: }
\begin{itemize}
\item[(a)] $\omega_0$ is a simple eigenvalue with a strictly positive eigenvector.

\item[(b)] $e^{-t \Phi}$ is positivity improving for all $t>0$. 

\end{itemize}

Therefore, having disposed of the preliminary steps given in the previous sections, we now only need to prove that the semi-group $e^{-t \Phi}$ is positivity improving for all $t>0$ since we know that $\Phi$ is a self-adjoint operator which is  bounded from below.
One can easily check that $e^{-t K_0}$ is a positivity improving semi-group for all $t>0$ due to the following theorem \cite{reedsimonv4}: 

\textit{Let $A$ be an operator on a complex Hilbert space with a distinguished complex conjugation and which obeys $A \geq c I$ for some $c>0$. Then, $e^{-t d\Gamma(A)}$ is positivity improving for all $t>0$, where $d\Gamma(A)$ is the second quantization of $A$.}

As a consequence of the above theorem,  the semi-group generated by the second quantization of this operator, $e^{-tK_0}$ is positivity improving for all $t>0$ since $-{1 \over 2m} \nabla_{g}^{2} +m I \geq m I $. Another way of showing this is based on the idea that the semi-group generated by $K_0$ can be expressed in terms of the heat kernel $K_t(x,y)$ which is strictly positive as long as $t>0$.

Before investigating the positivity property of the semi-group generated by the remaining part of the principal operator,  we make the following observation now.
\textit{If $e^{-tA} $ is positivity improving and $e^{-tB}$ is positivity preserving for all $t>0$, then 
the product 
$e^{-sA}(e^{-tB})$ is positivity improving for all $t>0$.}

The second real-valued kinetic operator $K_1$ is a positive self-adjoint operator on the domain 
$D(K_1)= \{ \psi \in L^2(M, d \mu)| \int k_{1}^{2} |\psi|^2 \; d \mu < \infty\}$.
One can now show that $\psi \in D(K_1)$ if $\psi \in D(H_0)=D(K_0)$. This can be seen by using the spectral theorem and the fact that $K_1$ is a function of the positive self-adjoint operator $H_0$.
In other words, the domain of  $K_1$ contains the domain of $H_0$, i.e., $D(K_1) \supset  D(H_0)$. Also, one can show that quadratic form domain of  $K_1$ includes the quadratic form domain of $H_0$. This can be seen easily now, since, as shown in Sec. \ref{Existence of the Ground State},  $K_0+K_1$ is  a relatively compact perturbation of $K_0$.  Both of them are positive operators, so that they naturally define positive quadratic forms.
The operators $K_0$ and $K_1$, defined via the same spectral measure, obviously commute.

We will now show that $e^{-t K_1}$ is positivity improving for all $t>0$.
The compactness of a Riemannian manifold implies that
it is complete as a Riemannian manifold and it has a Ricci curvature tensor bounded from below, i.e.,  
$Ric(\cdot, \cdot)\geq -\kappa g(\cdot,\cdot)$. 
As a result of the theorem proven by J. Cheeger and S.-T. Yau \cite{CheegerYau}, the heat kernel has the following lower bound
\be
K_t(x,y) \geq K_{t}^{\kappa}(d_g(x,y)) \;, \label{lowerbound heatkernel}
\ee
where $K_{t}^{\kappa}$ is the heat kernel for the simply connected complete Riemannian manifold of constant
sectional curvature $-\kappa$ and $d_g(x,y)$ refers to the geodesic distance on the manifold ${\cal M}$. In particular, we choose $K_{t}^{\kappa}(d_g(x,y))$ as the heat kernel for the Hyperbolic manifold $\mathbb{H}^{D}$. 
In three dimensions, since we have an explicit expression of the heat kernel \cite{Grigoryan Heat Book},  the lower bound to it is simply 
\begin{equation}
   K_t(a,a)\geq {e^{-\kappa t/2m}\over (4\pi t/2m)^{3/2}} \;.
\end{equation}
We add and subtract this lower bound to the heat kernel in $K_1$ and then split it to the following two parts $K_r$ and $K_p$, defined as, 
\begin{eqnarray}
  K_r(E)&=&\lambda^2 \int^\infty_0  \underbrace{\left(K_t(a,a)-{e^{-\kappa t/2m}\over (4\pi t/2m)^{3/2}}\right)}_{\geq 0} \, \left(e^{-t(m-\mu)}-e^{-t(H_0+m-E)}\right) \; dt \nonumber\\
  K_p(E)&=&\lambda^2 \int_0^\infty  {e^{-\kappa t/2m}\over (4\pi t/2m)^{3/2}} \left(e^{-t(m-\mu)}-e^{-t(H_0+m-E)} \right) \; d t \nonumber\\
            &=&{\lambda^2\over (4\pi)^{3/2}}(2m)^{3/2}  \left[\left(H_0-E+m+{\kappa \over 2m}\right)^{1/2} - \left(m+{\kappa \over 2m}-\mu \right)^{1/2} \right] \;,
\end{eqnarray}
where $E$ is real and $E<nm+\mu$. Since all the projections in their associated projection - valued measures of $K_r$ and $K_p$ commute, we have
\be e^{-t K_1(E)}=e^{-t K_r(E)-t K_p(E)}=e^{-t K_p(E)}e^{-t K_r(E)} \;.
\ee 
We will first prove that the semi-group generated by $K_p(E)$ is actually {\it positivity improving} for all $t>0$ in three dimensions:
\begin{equation} 
e^{-tK_p(E)}=e^{t \lambda^2 C \sqrt{(m+{\kappa \over 2m}-\mu)}} e^{-t\lambda^2 C\sqrt{H_0+{\kappa \over 2m}+m-E}} \;,
\end{equation}
where $C= \left( {2m \over 4\pi} \right)^{3/2}$. If we apply the subordination identity to the last piece;
\begin{equation}
 e^{-t\lambda^2 C\sqrt{H_0+m+ {\kappa \over 2m}-E}}=
t{C\lambda^2\over 2\sqrt{\pi}}  
\int_0^\infty  {e^{-t^2C^2\lambda^4/4u} e^{-uH_0} e^{-u({\kappa \over 2m}+m-E)} \over u^{3/2}} \; d u \;,
\end{equation}
which is {\it explicitly positivity improving} for all $t>0$ since $e^{-t H_0}$ is so and everything else is positive in the integration.
We will now remark that the remaining part, which is given by $K_r(E)$ is actually positivity preserving for all $t>0$. This can be proven by the use of Beurling-Deny criteria \cite{reedsimonv4}:

\textit{
Let $H$ be a self-adjoint positive operator on $L^2$. The quadratic form $(\psi, H \psi)$ is extended to all of $L^2$ by setting it equal to infinity when $\psi \notin Q(H)$. Then,
the semi-group $e^{-t H}$ generated by a self-adjoint, positive operator $H$ is positivity preserving for all $t>0$ if $H$ satisfies the following condition for all $\psi$ in the Hilbert space:
\begin{equation}
\langle |\psi| , H |\psi| \rangle  \leq \langle \psi, H \psi \rangle \;.
\end{equation}}
In our case, this  condition for $K_r$ can be checked as follows, 
\begin{eqnarray}
   & & \langle |\psi|, K_r(E)|\psi|\rangle = \lambda^2 \langle |\psi|,\int_0^\infty  \underbrace{\left(K_t(a,a)-{e^{-\kappa t/2m}\over (4\pi t/2m)^{3/2}}\right)}_{\geq 0} \, \left (e^{-t(m-\mu)}-e^{-t(H_0+m-E)} \right)  |\psi|\rangle \; d t  \nonumber\\
&  & = \lambda^2 \int_0^\infty  \left( K_t(a,a)-{e^{-\kappa t/2m}\over (4\pi t/2m)^{3/2}}\right) \, \left(\langle |\psi|,|\psi|\rangle e^{-(m-\mu)t} -\langle e^{-t(H_0+m-E)/2}|\psi|, e^{-t(H_0+m-E)/2}|\psi|\rangle \right) \; d t \nonumber\\   & & \leq \lambda^2 \int_0^\infty  
\left(K_t(a,a)-{e^{-\kappa t/2m}\over (4\pi t/2m)^{3/2}}\right)
 \left(\langle \psi,\psi\rangle e^{-(m-\mu)t} -\langle |e^{-t(H_0+m-E)/2}\psi|, |e^{-t(H_0+m-E)/2}\psi|\rangle \right) \; d t \label{K_1 beurling deny1} \;,
\end{eqnarray}
where we have used the fact that $H_0$ is self-adjoint and $e^{-t (H_0 +m -E)/2}$ satisfies the Beurling-Deny criteria since it is positivity improving for all $t>0$. Then, by using the self-adjointness of  $H_0$ once more, the last line in the above equation becomes
\begin{eqnarray}
& & \lambda^2 \int_0^\infty \left(K_t(a,a)-{e^{-\kappa t/2m}\over (4\pi t/2m)^{3/2}}\right)  \left(\langle \psi,\psi\rangle e^{-(m-\mu)t} -\langle e^{-t(H_0+m-E)/2}\psi, e^{-t(H_0+m-E)/2}\psi\rangle \right) \; d t \nonumber\\
& & = \lambda^2 \int_0^\infty  \left(K_t(a,a)-{e^{-\kappa t/2m}\over (4\pi t/2m)^{3/2}}\right) \left(\langle \psi,\psi\rangle e^{-(m-\mu)t} -\langle \psi, e^{-t(H_0+m-E)} \psi\rangle \right)\; d t \nonumber\\
& & = \langle \psi, K_r(E)\psi\rangle \;.
\end{eqnarray}
This same condition in two dimensions requires more care, because the heat equation on $\mathbb{H}^2$ is given by an integral expression which is hard to estimate.
Davies and Mandouvalos \cite{Davies Mandouvalos} have obtained the sharp upper and lower bounds of the heat kernel on hyperbolic manifolds, which give rise to remarkable consequences for us. This lower bound of the heat kernel for $\mathbb{H}^2$ is given by 
\be
K_t(x,x) = \lim_{x \rightarrow y} K_t(x,y) \geq  {c \; e^{- \kappa t/8m} \over (t/2m) \left(1+ {\kappa t \over 2m}\right)^{1/2}} \geq {c \; e^{- 3 \kappa t/8m} \over (t/2m)} \;,
\label{lowerboundheatkernel2}\ee
since $(1 + \kappa t/2m)^{-1/2} \geq e^{- \kappa t/4m}$ for all $t \geq 0$ and here $c$ is a positive dimensionless constant.

Using the lower bound (\ref{lowerboundheatkernel2}), the same decomposition  of $K_1$ in a two dimensional compact Riemannian manifold  leads to the following expression for $K_p(E)$:
\begin{equation}
\lambda^2 c (2m) \ln \left({H_0+m+{3\kappa \over 8m}-E\over m-\mu + {3\kappa \over 8m}}\right) \;.
\end{equation}
The semi-group generated by this operator becomes,
\begin{equation}
e^{-t K_p(E)}=  \left({H_0+m+{3\kappa \over 8m}-E\over m-\mu + {3\kappa \over 8m}}\right)^{-\lambda^2 c t} \;,
\end{equation}
and  this can be written as 
\begin{equation}
 \left({H_0+m+{3\kappa \over 8m}-E\over m-\mu + {3\kappa \over 8m}}\right)^{-\lambda^2 c t}
= {1\over \Gamma(\lambda^2 c t)} \int_0^\infty   { e^{-s(H_0-E+m+{3\kappa \over 8m})/(m-\mu + {3\kappa \over 8m})}  \over s^{1-c \lambda^2 t}}\; d s \;,
\end{equation}
so that the semi-group $e^{-t K_p}$ in two dimensions is {\it positivity improving} for all $t>0$, as well.
Similar to the three dimensional case, Beurling - Deny criteria for $e^{-t K_r}$ in two dimensions can be easily checked, hence it is positivity preserving for all $t>0$. This leads to the conclusion that $e^{-t K_1}$ is positivity improving for all $t>0$.

Let us recall the Trotter - Kato product formula \cite{reedsimonv1}:

\textit{Let $A$ and $B$ be two self-adjoint operators which are bounded from below, and asume that  the  sum $A+B$ is self-adjoint on a common domain. Then we have 
\begin{equation} 
e^{-t (A+B)} =\underset{{N\to \infty}}{\slim} \Big(e^{-t A/N}e^{-t B/N}\Big)^N \;.
\end{equation} }
We will now apply this theorem to the principal operator $\Phi(E)=K_1(E)+K_0(E)+U(E)$ (note the change of ordering). We have proved in our previous work that the principal operator is a self adjoint operator for real $E$ \cite{existence}. 
We also note that $K_1(E)>0$ explicitly  for  $E < nm + \mu$ and thanks to the estimate of the bottom of the ground state (\ref{E bound}), we have $K_0(E)+U(E)>0$ as long as $E\leq E_*$ .
Moreover, we have also shown that the difference of the principal operator corresponding to the two different values of $E$ is bounded. Similarly, we have
\be ||U(E_1)-U(E_2)||< |E_1-E_2| \lambda^2 n \int_0^\infty  t \; K_{2t}(a,a) \; e^{-t n m} \; dt \;. \ee
Hence for all values of $E_* < E < nm+ \mu$, we write
\begin{eqnarray} & & 
  \left(f^{(n)}, \left[K_0(E)+U(E)\right] f^{(n)} \right) = \left(f^{(n)},  \left[ H_0-E_*+\mu +U(E_*)-E+E_* +U(E)-U(E_*) \right] f^{(n)} \right) \nonumber\\
& & \hspace{2cm} > \underbrace{ \left(f^{(n)},(K_{0}(E_*) + U(E_*))f^{(n)} \right)}_{>0}-  |E_*-E|\left( 1+n\lambda^2\int_0^\infty  t \; K_{2t}(a,a) e^{-t nm} \; dt \right) \;,
\end{eqnarray}
where $f^{(n)}$ is any $n$-particle state. It shows that for $E<nm+\mu$, $K_0(E)+U(E)$ is bounded from below.

Thus we may now apply the Trotter-Kato formula
\begin{equation}
 e^{-t\Phi(E)}=\underset{{N\to \infty}}{\slim} \Big( e^{-K_1(E)t/N}e^{-(K_0(E)+U(E))t/N}\Big)^N \;.
\end{equation} 
Note that here we may rewrite $U(E)$ in the following way
\begin{equation}
U(E)=-\lambda^2\int_0^\infty \phi^\dagger(K_t(a,.)) \; e^{-t(H_0-E+2m)} \; \phi(K_t(a,.)) \; d t \;.
\end{equation}
We will now assure that the semi-group $e^{-t(K_0(E)+U(E))}$ is positivity preserving for all $t>0$.
We again resort to the Beurling-Deny criteria and check this condition only for $U(E)$ since it is obviously true for $K_0(E)$-part;
\begin{equation}
\langle |\psi|, U(E)|\psi|\rangle=-\lambda^2 \int_0^\infty  \langle e^{-t(H_0-E+2m)/2} \phi_g(K_t(a,.))|\psi |, e^{-t(H_0-E+2m)/2} \phi_g(K_t(a,.))|\psi|\rangle \; d t  \;.
\end{equation}
Using the positivity of $K_t(a,x)$ and the fact that $e^{-t(H_0-E+2m)}$ generates a positivity improving semi-group for all $t>0$, one can now  check that 
\begin{equation}
   e^{-t(H_0-E+2m)/2} \phi_g(K_t(a,.))|\psi|\geq |e^{-t(H_0-E+2m)/2} \phi_g(K_t(a,.))\psi| \;.
 \end{equation}
Thanks to the minus sign in front, we have now,
\begin{eqnarray}
\langle |\psi|, U(E)|\psi|\rangle&\leq& -\lambda^2 \int_0^\infty  \langle |e^{-t(H_0-E+2m)/2} \phi_g(K_t(a,.))\psi|, |e^{-t(H_0-E+2m)/2} \phi_g(K_t(a,.))\psi| \rangle \; dt \nonumber \\
&\leq& -\lambda^2 \int_0^\infty \langle e^{-t(H_0-E+2m)/2} \phi_g(K_t(a,.))\psi, e^{-t(H_0-E+2m)/2} \phi_g(K_t(a,.))\psi \; dt  =\langle \psi, U(E)\psi\rangle \;.
\end{eqnarray}
Thus, we conclude that the semi-group $e^{-t(K_0(E)+U(E))/N}$  in the Trotter-Kato product formula is positivity preserving for all $t>0$.
Since the first factor in the  Trotter-Kato product expansion of $\Phi(E)$  is positivity improving and the second factor is positivity preserving for all $t>0$, their product $e^{-t K_1(E)/N}e^{-t (K_0(E)+U(E))/N}$ is positivity improving for all $t>0$. As a result, the principal operator $\Phi(E)$ generates a positivity improving semi-group for all $t>0$ due to the fact that the strong limit of the sequence of positivity improving operators are positivity improving for all $t>0$. Thus the eigenvalue corresponding to the bottom of the spectrum of principal operator is simple, i.e., nondegenerate and its associated eigenvector is strictly positive. 

Hence, due to Eq. (\ref{psi0_1}) and Eq. (\ref{psi_0_2}), the ground state wave function of the model is strictly positive so that  the ground state of the original model is nondegenerate.

\section{Acknowledments} 

The authors would like to thank J. Hoppe and B. T. Kaynak for their useful suggestions.

\end{document}